\documentclass[preprint]{ptephy_v1}

\preprintnumber{XXXX-XXXX} 
\usepackage{hyperref}
\usepackage[normalem]{ulem}

\newcommand{\WGy}[1]{{\color{green}#1}}

\renewcommand\sout{\bgroup \color{red} \ULdepth=-.5ex \ULset}

\begin{document}

\title{Studying the in-medium $\phi$ meson spectrum through kaons in proton-nucleus reactions}


\author{Gabor Balassa}
\affil{Department of Physics, Yonsei University, Seoul 03722, Korea\email{balassa.gabor@yonsei.ac.kr}}

\author{Kazuya Aoki}
\affil{KEK, High Energy Accelerator Research Organization, Tsukuba, Ibaraki 305-0801, Japan\email{kazuya.aoki@kek.jp}}

\author{Philipp Gubler}
\affil{Advanced Science Research Center, Japan Atomic Energy Agency, Tokai, Naka, Ibaraki 319-1195, Japan\email{philipp.gubler1@gmail.com}}

\author{Su Houng Lee}%
\affil{Department of Physics and Institute of Physics and Applied Physics, Yonsei University, Seoul 03722, Korea\email{suhoung@yonsei.ac.kr}}

\author{Hiroyuki Sako}
\affil{Advanced Science Research Center, Japan Atomic Energy Agency, Tokai, Naka, Ibaraki 319-1195, Japan\email{hiroyuki.sako@j-parc.jp}}

\author{Gy\"orgy Wolf}\affil{Institute for Particle and Nuclear Physics, Wigner Research Centre for Physics, H-1525 Budapest, Hungary\email{wolf.gyorgy@wigner.hun-ren.hu}}


\begin{abstract}%
Exploring the mass modifications of $\phi$ mesons in nuclei provides insights into the nature of strongly interacting matter. Specifically, $\phi$ meson mass shifts can be related to the in-medium modification of the strange quark condensate. Therefore, the partial restoration of chiral symmetry can be studied by observing the mass shifts through the decay channels $\phi \rightarrow e^+e^-$, and $\phi \rightarrow K^+K^-$. 
In this paper, we examine the possibility of observing the $\phi$ meson mass modifications of the $\phi$ mesons in 30 GeV proton-nucleus (C, Cu, Pb) collisions,  
to be studied at the J-PARC E88 experiment, through the kaonic decay channel, with the off-shell Budapest Boltzmann-Uehling-Uhlenbeck (BuBUU) transport model. By applying different mean fields to the kaons, we examine their effects on the invariant mass spectra.   
Our simulations suggest that, although different mean fields for the kaons do affect the spectrum, there is a common observable effect primarily driven by the $\phi$ mass shift.
However, due to the threshold associated with the two kaons, the signal we observe is quite different from the one expected in dilepton spectra. 
Therefore, to meaningfully constrain the mass shift, it will be useful to include both kaon and dilepton channels in the analysis of the experimental data.
\end{abstract}

\subjectindex{xxxx, xxx}

\maketitle

\section{Introduction}
\label{sec:Introduction}
The partial restoration of chiral symmetry at non-zero nuclear densities suggests that the values of the quark condensates will change, which is expected to have direct consequences for the spectral function of specific mesons, e.g. vector mesons, such as $\rho$ and $\phi$. Similarly, the change in, e.g., the gluon condensate at finite density will have consequences in the mass spectra of both the light and heavier charmonium or bottomonium states. By observing the changes in the particle properties, e.g., masses and widths, it could hence be possible to obtain insights into the properties of the strongly interacting matter, the origin of  their vacuum properties, and Lorentz-symmetry breaking effects such as momentum dependency of the mass shifts in the dense medium \cite{Hatsuda:1991ez,Lee:1997zta,Kim:2019ybi,Kim:2022eku}. A relatively clean environment (compared to nucleus-nucleus collisions) to observe such in-medium effects are offered by proton-nucleus collisions. We are going to observe the effects of the possible mass shifts with proton beam colliding with different target nuclei , e.g., C, Cu, and Pb, at a 30 GeV beam energy available at J-PARC. 
Here, we will consider the mass modification of the $\phi$ meson, which is one of the most promising candidates to observe such effects at moderate energy p+nucleus collisions that will be carried out at the J-PARC E16 \and E88 experiments~\cite{Aoki:2023qgl,Aoki:2024ood,Sako:2024oxb}, due to its sensitivity to the $\langle \overline{s}s \rangle$ condensate at finite densities\cite{Hatsuda:1991ez,Kampfer:2002pj,Gubler:2014pta,Gubler:2015yna}.

Vector mesons are indeed promising candidates to experimentally observe such changes due to their decay into dilepton pairs, which, after their creation in the dense region, will propagate mostly unaffected by their surroundings. Therefore, at the end of the collision, the invariant mass spectrum of the dilepton pairs could give us insight into the density of the region in which it was created \cite{KEK-PS-E325:2005wbm}.
One drawback of using such dilepton pairs is their very small branching fractions. Hence, a large number of events are necessary to observe the effects of the mass modifications. To overcome this problem, the hadronic decay channel $\phi \rightarrow K^+ K^-$ provides a suitable alternative for observing the $\phi$ meson mass shifts. It is not straightforward, however, what kind of signal will be obtained, as the kaons will ``strongly" interact with the surrounding nucleons after they are created in the dense matter. Therefore, the final signal could very well be smeared out even with the much higher statistics compared to the dilepton case. It is hence important to estimate the possible yields and shapes of the invariant mass spectrum of the kaon pairs to be able to understand the data that will be obtained, for instance, at the J-PARC E88 experiment \cite{Sako:2024oxb} with a 30 GeV proton beam colliding with different nuclear targets.

Previous studies have been carried out at 12 GeV bombarding energies that correspond to the KEK-E325 experiment, in which the $\phi$ meson spectral modification 
was analyzed through dileptonic and kaonic decays. While the interpretation of the $K^+K^-$ data was not clear \cite{E325:2006ioe}, 
the original analysis of the experimental dilepton data suggested a small $~34$ MeV negative mass shift \cite{KEK-PS-E325:2005wbm}. However, the statistical significance of the data was not enough to draw firm conclusions.

On the theoretical side, in-medium properties of hadrons are studied through off-shell transport codes, which take into consideration the full nonequilibrium evolution of the collisional system and can incorporate the mass and width changes by simultaneously respecting energy conservation. In \cite{Wolf:Charm-2018}, a Boltzmann-Uehling-Uhenbeck (BUU) type off-shell transport code (BuBUU, Budapest BUU) was used to describe the mass shifts of different low-lying charmonium states through their dilepton spectra, in which the value of the mass shifts could be related to the value of the gluon condensate at normal nuclear density. A similar approach that is frequently used is based on the Parton Hadron String Dynamics (PHSD) code \cite{Cassing:2008sv,Cassing:2009vt}, which is another off-shell transport method. In Refs.\,\cite{Gubler:2024ovg,PSE325:2025iku}, PHSD was employed to describe the mass shifts of the $\phi$ mesons through their dilepton decays, which was the prime objective of the KEK-E325 experiment, where the dilepton spectrum from $\phi \rightarrow e^+ e^-$ decays are observed in p+A collisions at 12 GeV bombarding energies. While the results of the experiment indeed show some slight modification corresponding to a small negative mass shift, the statistical significance of the data was hardly enough to draw firm conclusions. In the simulations made by the PHSD code in \cite{Gubler:2024ovg}, the dilepton spectrum is compared by applying different mass shifts and width changes, each showing possible observable signals, especially for larger nuclei. However, due to the possibly small mass shifts, the signal could be smeared out in a real experiment, as was indeed shown in \cite{PSE325:2025iku}.For real experiments, observing a two-peaked structure of the $\phi$ meson—as proposed in \cite{Ko-Seibert-1994} for detecting the deconfinement transition in heavy-ion collisions and also predicted in the case of $\psi(3686)$ in \cite{Wolf:Charm-2018}—would be desirable. This would, however, necessitate a much cleaner signal with larger mass shifts and a small background that is not always achievable.
Here, we will focus on the $\phi \rightarrow K^+ K^-$ channel to be measured at the future J-PARC E88 experiment \cite{Sako:2024oxb} by simulating p+nucleus collisions at 30 GeV bombarding energies with C, Cu, and Pb targets, using the off-shell BuBUU transport code discussed in Ref.\cite{Almasi-Wolf-2015}. To study
the effects of the interactions between the kaons and the nucleons that could distort the final state invariant mass spectrum, four different mean fields describing the $KN$ and $\bar{K}N$ interaction will be applied.  

The paper is organized as follows. 
In Sec.\,\ref{sec:1}, the main ingredients of the BuBUU transport code are described by focusing on the parts that are necessary to understand the later results. After the technical description, in Sec.\,\ref{sec:2}, the effects of the different kaon mean fields will be compared through the invariant mass spectrum of the kaon pairs, while in Sec.\,\ref{sec:3}, the results for the mass shifts in p+C, p+Cu, and p+Pb collisions at 30 GeV bombarding energies are compared by using the most up-to-date mean fields. Finally, in Sec.\,\ref{sec:Conclusions}, the main conclusions are drawn, and possible future developments that could help in understanding the forthcoming experiments will be discussed.

\section{Description of the BuBUU transport model}
\label{sec:1}
The measured data of a full nonequilibrium dynamical system that is formed during heavy ion collisions is usually very hard to interpret without appropriate numerical methods. One widely used approach is the Budapest Boltzmann-Uehling-Uhlenbeck type transport model, BuBUU \cite{Wolf-1990,Wolf-1993,Teis-1996,Wolf-2012,Almasi-Wolf-2015}, which is able to follow the dynamical evolution of the interacting, nonequilibrium system. There are several transport approaches available in the field of high-energy physics, each starting from some common dynamical equations. There are differences in the implementation of these equations, e.g., the way the 2-body collisions are treated, the mean fields applied to the different particles, or how the Pauli exclusion principle is taken into consideration. At moderate energies up to a few tens of AGeV, the partonic degrees of freedom are not necessary for an accurate description of the system. However, at higher energies (100 GeV to TeV range), where, e.g., the Quark Gluon Plasma could be formed, the partonic side also has to be included to be able to correctly assess the interactions between partons. Some of the codes used nowadays are capable of including off-shell propagation of particles, 
and can hence treat the mass and/or width modifications by respecting the energy conservation at relativistic energies.
 
To implement such simulations, we use the off-shell BuBUU transport code that was developed by some of us and has been used to solve and understand various problems in the past \cite{Wolf-2012}. Most recently it was applied to describing single and double hypernuclei production in heavy ion collisions \cite{Balassa-2023}. Regarding the off-shell propagation of particles, it was used to study the mass shifts of low-lying charmonium states in antiproton-induced reactions \cite{Wolf:Charm-2018}, and to compute the corresponding expectations from the invariant mass spectra of the dilepton pairs.  
Similarly, in this work, we simulate
the off-shell propagation of the $\phi$ mesons, but instead of examining the invariant mass spectra of the dileptons, 
we focus on the $K^+ K^-$ pairs to investigate what invariant mass spectrum could be observed in proton-induced reactions. 
In this section, we will give a short overview of the BuBUU transport approach used in this work.

In the energy range and type of collisions we are interested in, the partonic degrees of freedom can be safely neglected.
Therefore, we only consider hadronic degrees of freedom, including protons, neutrons, most of the nucleon $N$ resonances, the delta $\Delta$ resonances, the $\Lambda$ and $\Sigma$ baryons, the $\pi$, $\rho$, $\omega$, $\sigma$, $\eta$, and $K$ mesons, the $J/\Psi$, $\Psi(3686)$, $\Psi(3770)$ charmonium states, and the $D$ mesons. Apart from the usual Coulombic force acting on the charged particles, a momentum-dependent mean field is applied to the nucleons, specifically:
\begin{equation}
\label{eq:1}
U_N = A\frac{\rho}{\rho_0} + B \Big(\frac{\rho}{\rho_0}\Big)^\tau + 2\frac{C}{\rho_0}\int d^3p' \frac{f(\vec{r},\vec{p'})}{1+ \Big( \frac{\vec{p}-\vec{p'}}{\Lambda}\Big)^2},
\end{equation}
where $\rho_0$ is the normal nuclear density, $f(\vec{r},\vec{p})$ is the 6-dimensional phase space density, while A, B, and $\tau$, are free parameters that are fitted to the ground state properties of the nuclear medium. Furthermore, C and $\Lambda$ are fitted to the nuclear optical potential extracted from proton-nucleus collisions \cite{Teis-1996}. The mean field takes into account all the possible interactions between the nucleons and their surroundings, and one of its main effect is to change the effective masses of particles depending on the density they are propagating through. 

In this work we are mainly interested in the propagation of the kaons that are created through $\phi$ meson decays. Thus, the interaction between kaons and their surrounding nucleons will be particularly important. To take this into account, we applied several density- and momentum-dependent Ans{\"a}tze for the mean field that acts on kaons. 
Such mean field should be able to reproduce the kaon yields in previous heavy-ion reactions and the properties of kaonic atoms \cite{Friedman:1993cu}. It has to be noted, however, that even though the used kaon potentials are able to describe some experimental results, their explicit form is not 
uniquely determined. With this in mind, in the next section, we will compare the kaon invariant mass spectra in p+Cu collisions at 30 GeV incident energy under different mean fields and examine the differences that occur when we include the absorption of $K^-$ on nucleons, which would otherwise need to be accounted for through an imaginary part in the mean field potentials.

Apart from the mean fields and the Coulomb force, the elastic and inelastic collisions of the participating particles are also taken into account. Different codes use different approaches, and it is not straightforward to include 2 or many body collisions in a relativistically invariant way. The collisional criteria we applied for 2-body collisions is a Lorentz covariant description that minimizes causality violations that arise in, e.g., space-like events and causes non-invariant time ordering by imposing a causality condition for proper time intervals in the 2-body interactions. In \cite{Balassa-2023}, these collisional criteria 
have been extended to describe the yields and energy distributions of different fragments of the final state nucleons and used to estimate the yields of multiple hypernuclei in central Au+Au collisions up to 20 A GeV energies. The BuBUU code takes into account many elastic and inelastic reactions of the baryons and mesons, as described in \cite{Wolf-1997-HIPH}. As we deal in this work with $\phi$ mesons generated in proton+nucleus collisons, the most important elementary reactions are the creation, absorption, and other elastic and inelastic interactions of the $\phi$, $K^+$, and $K^-$ mesons. In contrast to heavy-ion collisions, where apart from the nucleons many $\Delta$ resonances and $\pi$ mesons are created, here the most abundant particles will be the nucleons. Thus, the most dominant channel to create the $\phi$ mesons is the $N+N\rightarrow \phi + X$ inclusive reaction, while near threshold the $\pi+N \rightarrow \phi + X$ reaction could also give some non-negligible contributions to the total yields. For the inclusive cross sections, we used the vacuum cross sections in \cite{Song_2022} that are fitted to experimental data that are collected in \cite{Baldini_88, Moskal_2002}. 

At high enough energies (more than 10 GeV), it is advisable to use the inclusive channels for creating the $\phi$ mesons needed to give reliable estimates of the total yields. At a few GeV bombarding energies, the $N+N \rightarrow N+N+\phi$ and $\phi + N \rightarrow \phi + N$ channels will, however, be sufficient. Apart from the dominant ones, the following channels are also included: $N+\Delta \rightarrow N+N+\phi$, $\Delta+\Delta \rightarrow N+N+\phi$, $N+\pi \rightarrow N + \phi$, $\Delta + \pi \rightarrow N + \phi$, $N+\rho \rightarrow N + \phi$, $\Delta + \rho \rightarrow N + \phi$, $\pi + \pi \rightarrow \phi + \pi$, $N+\omega \rightarrow N + \phi$, $\Delta+\omega \rightarrow N + \phi$, $N+\eta \rightarrow N + \phi$, $\Delta+\eta \rightarrow N + \phi$ 
\cite{Chung_Ko:1997-phielement,Barz-Wolf:2002-phiprod}. These, however, give only negligible contributions to the total yields in proton induced reactions. As we also create kaons through the decay of the $\phi$ mesons and from $NN$ and $N\pi$ reactions, it is possible that the $\phi$ mesons are created through the $K^+ + K^- \rightarrow \phi$ channels, which is also taken into consideration through a Breit-Wigner parametrization \cite{Chung:1998ev}.

The created $\phi$ mesons can also interact with their surrounding nucleons, which is taken into account by including the following parametrization to the $\phi N$ elastic cross section \cite{Song_2022, Golubeva_1997}, 
\begin{equation}
\label{eq:3}
\sigma_{\phi N \rightarrow \phi N}^{elastic} = a + \frac{b}{p_{lab}},
\end{equation}
where $p_{lab}$ is the momentum of the $\phi$ meson in the nucleon rest frame, while $a=10$ [mb] and $b=10$ [mb$\cdot$GeV/c]. Apart from the elastic interactions, it is also possible that the $\phi$ meson is absorbed by the nucleons, thus lowering the final yields depending on the density of the region it travels through. The final yield is therefore a complex interplay between creation, interaction, absorption, and re-creation of the $\phi$ mesons that is not straightforward to estimate without carefully analyzing the evolution of the system. The absorption is taken into account by a similar parametrization as the elastic contribution \WGy{\cite{Song_2022, Golubeva_1997}}, namely
\begin{equation}
\label{eq:3}
\sigma_{\phi N\rightarrow X}^{absorption} = a + \frac{b}{p_{lab}},
\end{equation}
where the parameters are $a=5$ [mb], and $b=4.5$ [mb$\cdot$GeV/c]. 

Unstable particles with finite widths could furthermore decay into different final states during their evolution in the dense system, which is taken into account through the decay probability defined as
\begin{equation}
\label{eq:3}
P = 1- \exp\Big(-\frac{\Gamma \Delta t}{\hbar \gamma }\Big),
\end{equation}
where $\Gamma$ is the decay width of the resonance, $\gamma$ is the Lorentz factor, while $\Delta t$ is the timestep of the simulation.
In general $\Gamma$ can be energy dependent, which is the case, e.g., for the $\Delta(1232)$ resonance. For the $\phi$ meson in this work, we will consider the following momentum-dependent form:
\begin{equation}
\label{eq:gamma}
\Gamma =\Gamma_{PDG} \left(\frac{p}{p_{PDG}}\right)^3. 
\end{equation} 
Here, the PDG subscript refers to the standard decay momentum  calculated using the masses from the PDG. 
The justification of the inclusion of the momentum dependence comes from the fact that the $\phi$ meson mass is very close to the kaon pair threshold. Therefore, any mass modification of the kaons or the $\phi$ strongly modifies the width of the $\phi$. Since the $\phi \rightarrow K^+K^-$ is a p-wave process, the width of the $\phi$ is proportional to $p^3$ where $p$ is the kaon momentum in the rest frame of the $\phi$ meson.

The main decay channels for the $\phi$ mesons are the $\phi \rightarrow K^+K^-$ with a Br=$0.491$ branching ratio, the $\phi \rightarrow K^0_L K^0_S$ with Br=$0.339$, and the $\phi \rightarrow 3 \pi$ pionic decay channel with Br=$0.154$. The other hadronic channels are negligible. Among the non-hadronic final states, the only important one for us is the $\phi \rightarrow e^+e^-$ dilepton channel, which will be later compared to the invariant mass spectrum of the kaon pairs. 

The kaon pairs are created using their shifted masses according to the applied mean fields, with the corresponding four-momenta that are determined by the two-body decay kinematics. Due to the shifted masses of the kaons, the thresholds for the nucleon+kaon interactions are corrected by a simple shift in $\sqrt{s}$.
When the kaon pairs are created in the simulation, their ``mother" particle $\phi$ is destroyed, but the origin of the pair is recorded, which facilitates to collect the corresponding pairs at the end of the collisions. 

As mentioned before, the kaons are interacting with the surrounding nucleons through a specific mean field, which can affect their behavior 
depending on the density around them at some specific time interval. Apart from the mean field, the elastic scatterings of $K^+$ and $K^-$ with nucleons 
are taken into account. For expressing the related cross sections, we employ the functional forms
\begin{align}
\label{eq:crosssection_param}
f_1(a,b;p) &= a e^{-b\cdot p}, \\
f_2(a,b;p) &= a p^{-b}, \\
f_3(a,b,c;p) &= a e^{(p - b)^2/c},  
\end{align}
where $a$, $b$ and $c$ are parameters with appropriately chosen units and $p$ stands for a momentum variable. 
Specifially, we use
the following parametrizations for the cross sections \cite{Hirtz:2018ocl}. 
\begin{equation}
\label{eq:3}
\sigma_{K^+N \rightarrow K^+N} =
\begin{cases} 
    a^+_1 & p_{lab}<935 $ $ \text{MeV/c}, \\ 
    a^+_2 - f_1(b^+_2,c^+_2;p_{lab}) & 935 \leq p_{lab} < 2080 $ $ \text{MeV/c}, \\ 
    f_2(a^+_3,b^+_3;p_{lab}) & 2.08 \leq p_{lab} < 5.5 $ $ \text{GeV/c}, \\ 
    a^+_4 & 5.5 \leq p_{lab} < 30 $ $ \text{GeV/c}.
\end{cases}
\end{equation}
Here, $p_{lab}$ is the kaon momentum in the nucleon rest frame. The elastic cross section for the $K^-$ is then described as
\begin{align}
\label{eq:3.1}
\sigma_{K^-N \rightarrow K^-N}  =&  f_2(a^-_1,b^-_1;p_{lab}) + f_3(a^-_2,b^-_2,c^-_2;p_{lab}) \nonumber \\ 
& + f_3(a^-_3,b^-_3,c^-_3;p_{lab}) + f_3(a^-_4,b^-_4,c^-_4;p_{lab}),
\end{align}
where the $p_{lab}$ momentum is given in [GeV] and the cross sections are given in [mb]. The parameters for the $K^+N$, and $K^-N$ interactions are collected in Tab.~\ref{tab:1}. 
\begin{table}[h!]
\centering
\begin{tabular}{c c c} 
 \hline
  & $K^+ N$ & $K^- N$  \\  
 \hline\hline
 $a^{\pm}_1$ & 12.0 & 6.132  \\ 
 $b^{\pm}_1$ & - & 0.2437  \\
 $a^{\pm}_2$ & 17.4 & 12.98  \\ 
 $b^{\pm}_2$ & 3.0 & 0.9902  \\
 $c^{\pm}_2$ & 6.3$\cdot10^{-4}$ & 0.05558  \\ 
 $a^{\pm}_3$ & 835.0 & 2.928  \\ 
 $b^{\pm}_3$ & 0.64 & 1.649  \\
 $c^{\pm}_3$ & - & 0.772  \\ 
 $a^{\pm}_4$ & 3.36 & 564.3  \\ 
 $b^{\pm}_4$ & - & 0.9901  \\
 $c^{\pm}_4$ & - & 0.5995  \\ 
 \hline
\end{tabular}
\caption{Fitted parameters for $K^+ N$ and $K^- N$ interactions from \cite{Hirtz:2018ocl}. The parameters are defined such that, when the laboratory momentum $p_{lab}$ is expressed in [GeV], the resulting cross sections are given in [mb].}
\label{tab:1}
\end{table}

Regarding the absorption of the $K^-$ mesons, we have included the dominant inelastic channels from \cite{Hirtz:2018ocl}, namely the $\overline{K}N\rightarrow \Lambda \pi$, $\overline{K}N\rightarrow \Sigma \pi$, $\overline{K}N\rightarrow \overline{K}' N'$, $\overline{K}N\rightarrow \Sigma \pi \pi$.

Before discussing the effects of the different mean fields and final state interactions of the kaons, 
let us briefly remark on the off-shell propagation of particles in this work. In the transport simulations, the $\phi$ mesons are treated as off-shell particles. Therefore, they are propagated according to the off-shell transport equations described in detail in \cite{Cassing:1999wx,Cassing:1999mh,Cassing:2000ch}. 
This means that, instead of a localized Dirac delta distribution, we represent the particles through their spectral function given as the relativistic Breit-Wigner distribution,
\begin{equation}
A(M) = \mathcal{N} \frac{M^2\Gamma}{(M^2-m^2)^2 + M^2\Gamma^2},
\end{equation}
where $M$ is the invariant mass, $m$ is the in-medium mass of the particle, $\Gamma$ is the in-medium width of the particle, and $\mathcal{N}$ is a normalization factor. Here, the in-medium masses can depend on the local density $\rho$, while the in-medium widths can depend on the density, the invariant mass $M$ and even on the velocity in the local rest frame.

While the equations of motion are similar to the on-shell case, they in the off-shell treatment include the real and imaginary parts of the retarded self-energies of the particles, among which the real part should depend on the mass shift and can be parametrized as, 
\begin{equation}
\mathcal{R}\Sigma^{ret} = 2m\Delta m_0 \frac{\rho}{\rho_0},
\end{equation}
where $m$ is the mass of the particle, $\Delta m_0$ is the mass shift parameter, $\rho_0=0.168$ [fm$^{-3}$] is the normal nuclear density, and $\rho$ is the local density. The actual particle mass shift will be $\Delta m = \sqrt{m^2+\mathcal{R}\Sigma^{ret}}-m \approx \Delta m_0 \frac{\rho}{\rho_0}$. $\Delta m_0$ hence represents 
the mass shift at normal nuclear density and is the main parameter that should be constrained by examining the 
invariant mass spectra of e.g., dileptons or kaon pairs.
The imaginary part of the self-energy depends on the vacuum width and the interaction between the particles through collisional broadening as
\begin{equation}
\mathcal{I}\Sigma^{ret} = -m(\Gamma + \Delta \Gamma \frac{\rho}{\rho_0}),
\end{equation}
where $\Gamma$ is the vacuum width given by Eq.~(\ref{eq:gamma}), and $\Delta \Gamma$ represents the width change that incorporates, e.g., collisional broadening effects, which depend on the relevant collisional cross sections and the velocities of the particles.  

When a meson is generated at finite density, its mass is determined randomly, sampled from the in-medium spectral function. During propagation, its mass is subsequently modified according to the real part of the self-energy. 
Similarly, due to the interactions with the medium, the width can also change according to the imaginary part of the self-energy. The corresponding $\Delta m_0$ and $\Delta \Gamma$ parameters either have to be fitted to experiments or have to be obtained by other means, e.g., theoretical QCD 
sum rule calculations \cite{Kim:2022eku}.

\section{Effect of the kaonic mean fields}
\label{sec:2}
In this section, the effects of the mean fields applied to the kaons and their interaction with the surrounding medium will be analyzed in $30$ GeV proton+Cu reactions. The aim of these simulations is to get a better understanding of how the kaons interact with the medium after they are generated from $\phi$ meson decays and how these interactions modify the invariant mass spectra of the corresponding kaon pairs. In contrast to dileptons, which after 
their creation mostly leave the dense matter freely, the kaons could be affected by the density- and momentum-dependent mean field that is generated by the collective interactions between kaons, nucleons, and other hadrons. 
Furthermore, elastic and inelastic collisions of kaons—primarily with nucleons—and the absorption of $K^-$  mesons could severely distort the spectrum, as the pair originating from the original $\phi$ meson would no longer exist in such cases.
The mean fields also depend on the electromagnetic forces due to the positive and negative charges of the kaons. To summarize, it is not straightforward to interpret a measured signal due to the many collective effects that are involved in determining the kaon pair invariant mass spectra. It is, therefore, indispensable to estimate each of their effects independently, for which the BuBUU transport code gives an appropriate tool, 
as we can simply turn on and off the effects of interest. 

To examine the effects of the different mean fields, we applied four different models that are shown in Tab.~\ref{tab:2}. 

\begin{table}[h!]
\centering
\begin{tabular}{c c p{6.5cm} p{5cm}} 
 \hline
 \textbf{Model} &  & \textbf{Mean Field Expression} & \textbf{Parameter Values} \\
 \hline\hline
 M1 & $K^+$ & $U_{K^+} = 0$ & -- \\
    & $K^-$ & $U_{K^-} = 0$ & -- \\
 \hline
 M2 & $K^+$ & $U_{K^+} = a_2 \cdot \rho$ & $a_2 = 0.167\;[\text{GeV} \cdot \text{fm}^3]$ \\
    & $K^-$ & $U_{K^-} = -a_2 \cdot \rho$  \\
 \hline
 M3 & $K^+$ & $U_{K^+} = a_3 \cdot \rho$ & $a_3 = 0.167\;[\text{GeV} \cdot \text{fm}^3]$ \\
    & $K^-$ & $U_{K^-} = -\rho (b_3 + c_3 \cdot e^{-d_3 \cdot p_K})$ & $b_3 = 0.341\;[\text{GeV} \cdot \text{fm}^3]$, \newline $c_3 = 0.823\;[\text{GeV} \cdot \text{fm}^3]$,\newline $d_3 = 2.5\;[\text{GeV}^{-1}]$ \\
 \hline
 M4 & $K^+,K^-$ & Density and momentum-dependent mean fields extracted from Refs.\cite{Tolos:2006ny,Tolos:2008di} \\
 \hline
\end{tabular}
\caption{Mean field models for $K^+$, and $K^-$. To obtain the potentials in [GeV] units, the kaon momenta $p_K$ should be given in [GeV], while the densities should be given in [fm$^{-3}$].}
\label{tab:2}
\end{table}
 
\begin{figure}[!h]
\centering\includegraphics[width=6in]{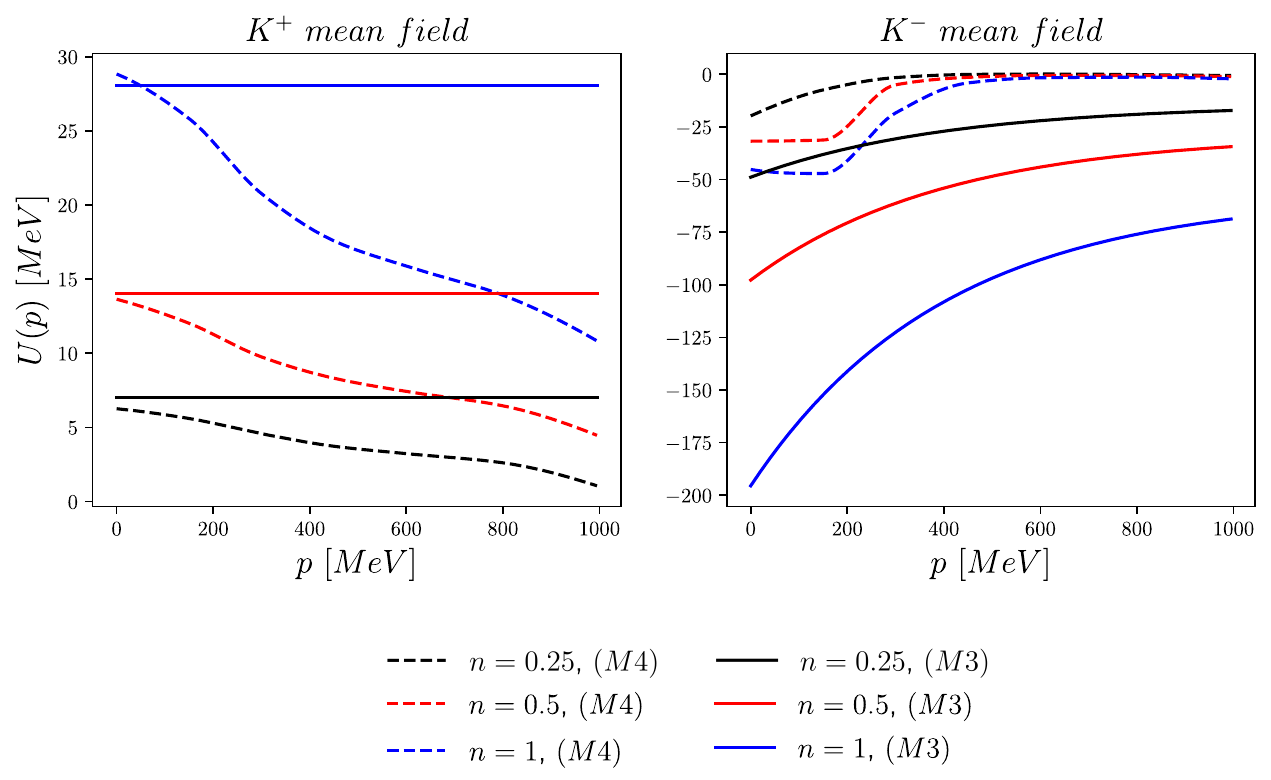}
\caption{Comparison of the density and momentum dependence of the (M3) and (M4) mean fields. The dashed lines show the results for the (M4), while the solid lines correspond to the (M3) potentials. Here, $n=\rho/\rho_0$ is the normalized density, where $\rho_0=0.168$ [fm$^{-3}$].}
\label{fig:kp_km}
\end{figure}

The first option (M1) acts as a reference for what to expect if no mean field is applied, while the second (M2) mean field is 
a non-realistic reference for what happens when the mean fields that act on the differently charged kaons are of the 
same magnitude but are repulsive (attractive) for $K^+$ ($K^-$). 
The third scenario (M3) is more realistic and is the result of an actual fit given in \cite{Sibirtsev:1998vz}, with different momentum dependences for $K^+$ and $K^-$ potentials.
The (M4) mean field is extracted via a self-consistent chiral unitary approach from in-medium meson-baryon amplitudes, giving a density- and momentum-dependent mean field for both kaons and antikaons. In Fig~\ref{fig:kp_km}, a comparison of the momentum dependent mean fields is shown for the two most realistic cases (M3) and (M4) at densities $\rho=0.25\rho_0$, $0.5\rho_0$, and $\rho_0$, where $\rho_0=0.168$ [fm$^{-3}$] is the normal nuclear density.

The comparison of the (M3) and (M4) mean fields shows a generally different behavior for the $K^+$ and $K^-$ at finite density. In the $K^+$ case with the (M3) potential, there is no momentum dependence, while in the (M4) case, there is a decreasing trend with higher momenta. The density dependence is similar in both cases. For $K^-$, the difference between the mean fields is more severe as the (M3) potential has much larger (negative) values at all momenta for the same densities. In the following, we will study how the invariant mass spectra of the final state kaons change according to the different potentials. 

\begin{figure}[!h]
\centering\includegraphics[width=4.5in]{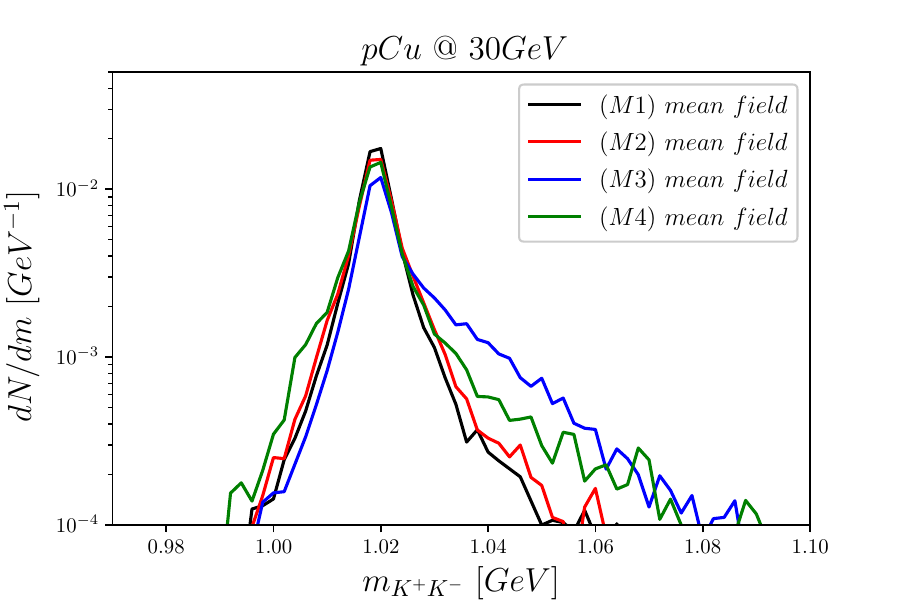}
\caption{Comparison of the invariant mass spectra of kaon pairs in p+Cu reactions at 30 GeV bombarding energies with different kaonic mean fields without final state interactions of kaons.}
\label{fig:test1}
\end{figure}

First, let us neglect the elastic and inelastic scattering and absorption processes of the created kaons and only consider the effects of the 
mean fields. The results for p+Cu collisions at $30$ GeV bombarding energies can be seen in Fig.~\ref{fig:test1}, where the $K^+ K^-$ invariant mass spectrum is shown. 
It is clear that the effects of the mean fields are indeed  significant, especially when the momentum dependence of the mean fields for the $K^+$ and $K^-$ are different, in which case the spectrum is deformed in an asymmetric manner. This deformation is caused by the interactions of the created kaons with the surrounding nucleons through their evolution in the dense matter, that is taken into consideration through the applied mean fields.
In contrast, when the mean fields are equal but with different signs for both kaons, this asymmetry is not expected to appear, which is indeed what we see in Fig.~\ref{fig:test1}.

\begin{figure}[!h]
\centering\includegraphics[width=4.5in]{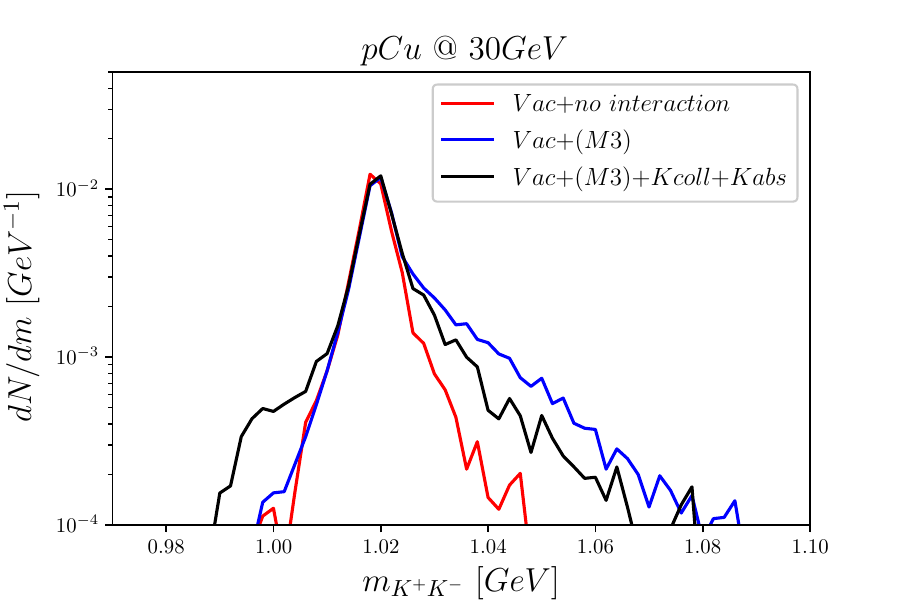}
\caption{Effects of the scatterings and absorption of kaons on the invariant mass spectrum when applying the (M3) mean field in p+Cu collisions at 30 GeV.}
\label{fig:test2}
\end{figure}

A more realistic scenario is obtained by additionally including the elastic scatterings of both $K^+$ and $K^-$, along with the absorption of $K^{-}$ particles during their evolution in dense matter. In Fig.~\ref{fig:test2}, the results for 30 GeV p+Cu collisions, with the (M3) mean field, are shown together with the previously shown (only mean field) case. 
The results indicate that including the final state interactions, an approximately symmetric structure of the spectrum is restored, similar to what was seen for the (M1) and (M2) mean fields in Fig.~\ref{fig:test1}. 

In the following, we will use the (M3) and (M4) mean fields, including the elastic and inelastic scatterings of kaons, to estimate the invariant mass spectra of the kaon pairs in proton-induced reactions. We will specifically include a finite negative mass shift for the $\phi$ mesons and compare the results for different target nuclei, namely C, Cu, and Pb at $30$ GeV bombarding energies.

\section{Results}
\label{sec:3}
\begin{figure}[h]
    \centering
    \includegraphics[width=0.55\textwidth]{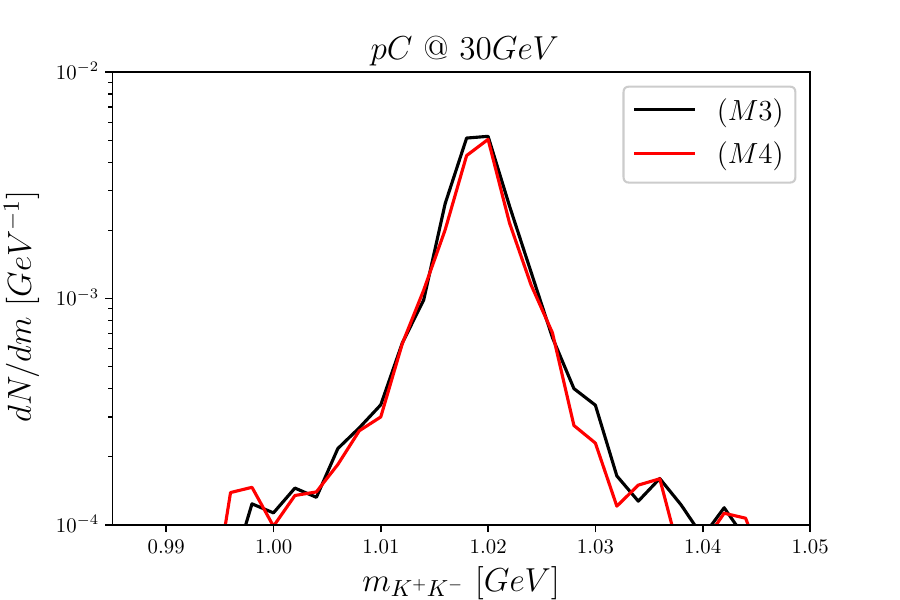}
    \includegraphics[width=0.55\textwidth]{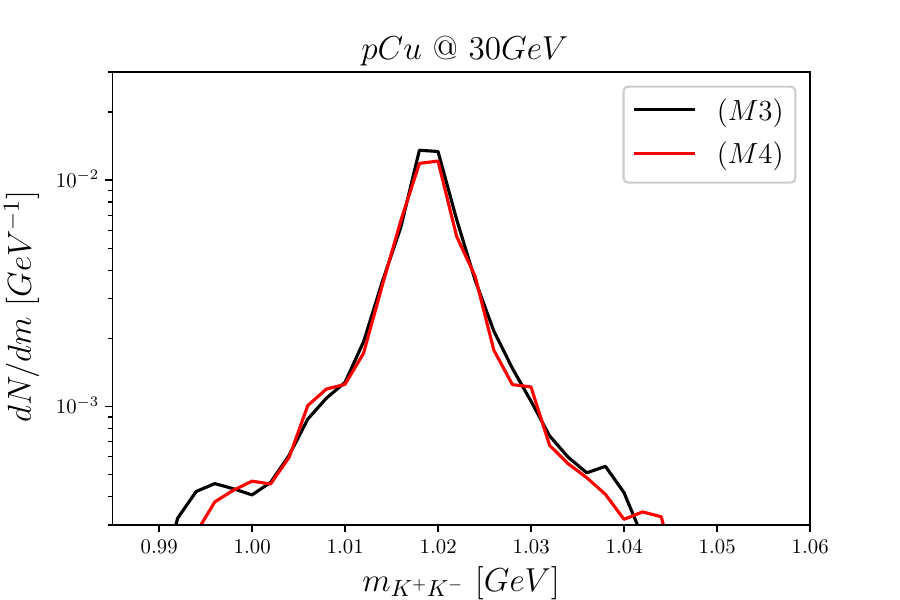}
    \includegraphics[width=0.55\textwidth]{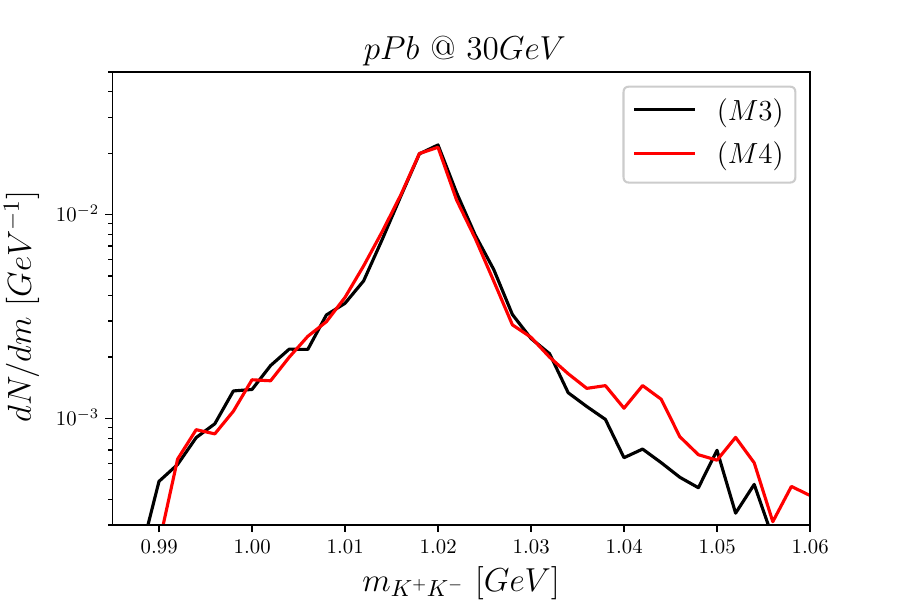}
    \caption{Comparison of the (M3), and (M4) mean fields at C, Cu, and Pb targets, including all of the considered final state interactions of the kaons.}
    \label{fig:plot31}
\end{figure}

In this section, we consider a finite, density dependent mass shift for the $\phi$ mesons and compare the respective
kaon invariant mass spectra with results obtained with a constant mass. Throughout, the most 
realistic (M3), and (M4) mean fields are applied to the kaons, for which the $K^+$ feels a small, density-dependent, and positive potential, while for the $K^-$ a larger, momentum- and density-dependent potential acts during their evolution in the dense matter. 
The $\phi$ meson mass shift is linear in density and is set to $\Delta m_{0}=-34$ MeV at normal nuclear matter density, according to the E325 experiment \cite{KEK-PS-E325:2005wbm}. To study pure mass shift contributions, we do not include any additional broadening effects, therefore, fixing the $\Delta \Gamma$ parameter in the imaginary part of the self-energy to $0$. To observe the effects caused by the different sizes of the targets, we compared the results using C, Cu, and Pb targets with proton beams at 30 GeV bombarding energies. We expect that the size of the system plays a crucial role in determining the shape of the kaon pair invariant mass distribution, due to the fact that in a larger system the mesons are propagating through a larger dense region before they reach the vacuum. In Fig.~\ref{fig:plot31}, the invariant mass spectra of the kaon pairs without mass modification $(\Delta m_0=0)$, and by including all the possible final state interactions of the kaons, are compared using the $(M2)$, $(M3)$ and $(M4)$ mean fields.

\begin{figure}[!h]
\centering\includegraphics[width=4.5in]{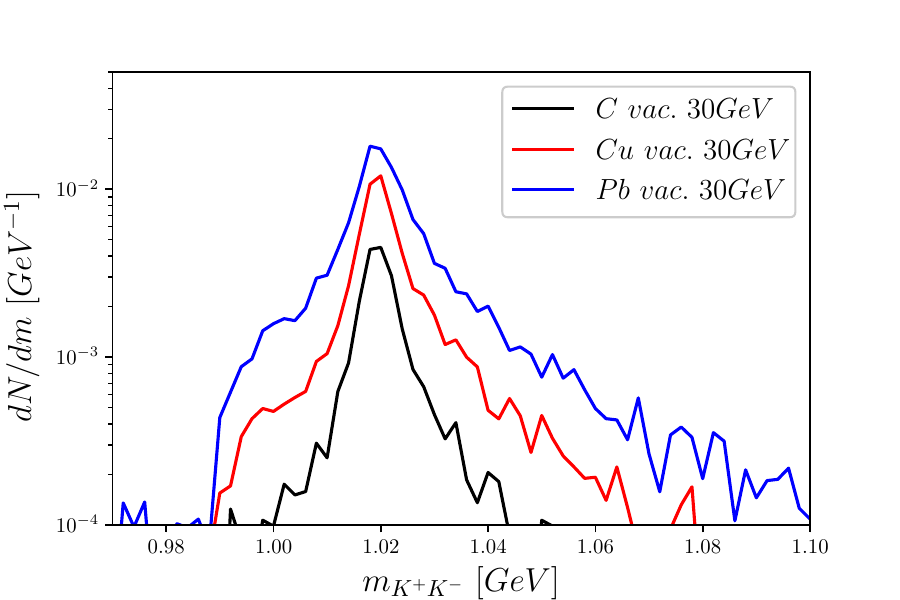}
\caption{Invariant mass spectra of K meson pairs without mass modifications of the $\phi$ mesons for C, Cu, and Pb targets in proton induced reactions at $30$ GeV bombarding energies by using the (M3) mean fields.}
\label{fig:32}
\end{figure}

The results indeed suggest that the system size plays an important role for the final shape of the invariant mass spectra. Due to the asymmetric nature of the (M3) mean field, a small but still observable deviation from the symmetric mean field arises for larger targets even with the inclusion of the final state kaon interactions, which was seen before (see Fig.~\ref{fig:test2}). For smaller targets, this deviation is less visible. In a real experiment, the expected yield of kaon pairs can also be a deciding factor for selecting targets. Therefore, in Fig.~\ref{fig:32}, the invariant mass spectra for the 3 targets are compared to each other by using the (M3) potential.
As expected, the yields are larger for larger targets, which is a direct consequence of the larger number of elementary $NN$ interactions that can create $\phi$ mesons. Note, however, that larger targets lead to more final state interactions, and thus a higher likelihood of absorption of the $\phi$ mesons and $K^-$ mesons, which can subsequently reduce the yields, which are hence a complex interplay of these effects. 

\begin{figure}[h]
    \centering
        \includegraphics[width=4.5in]{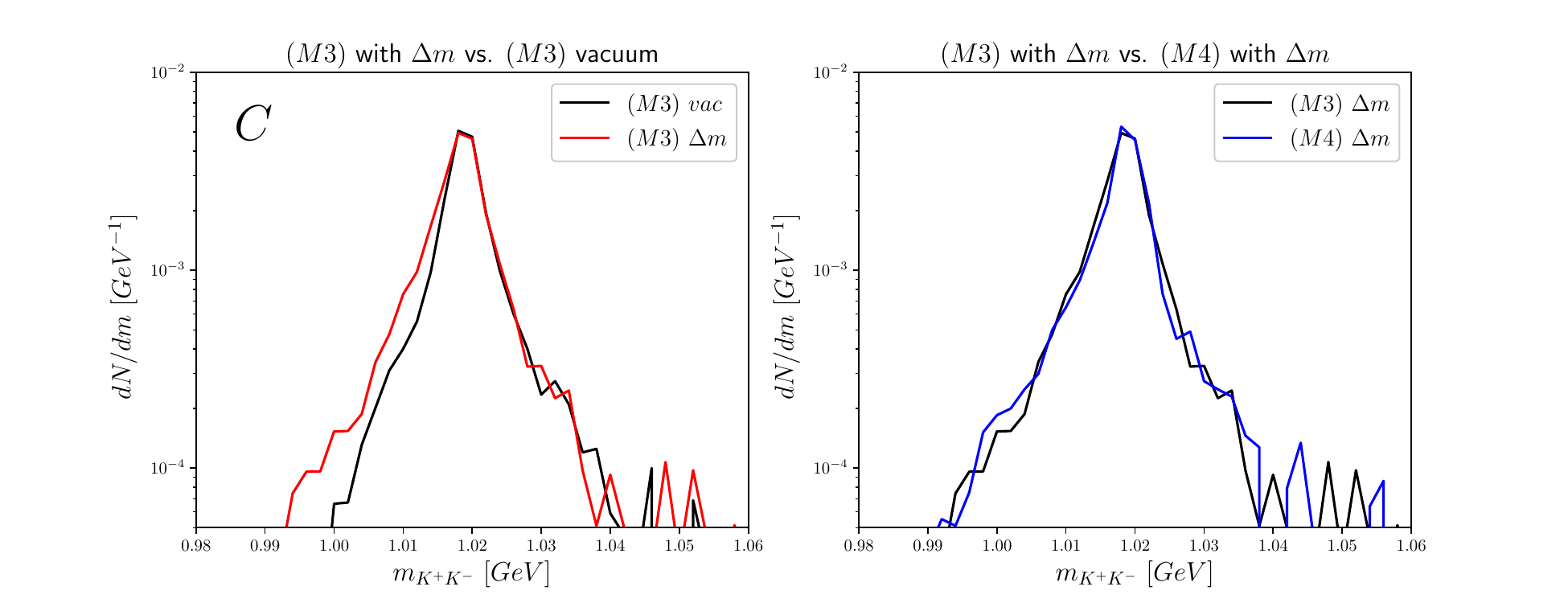}
        \includegraphics[width=4.5in]{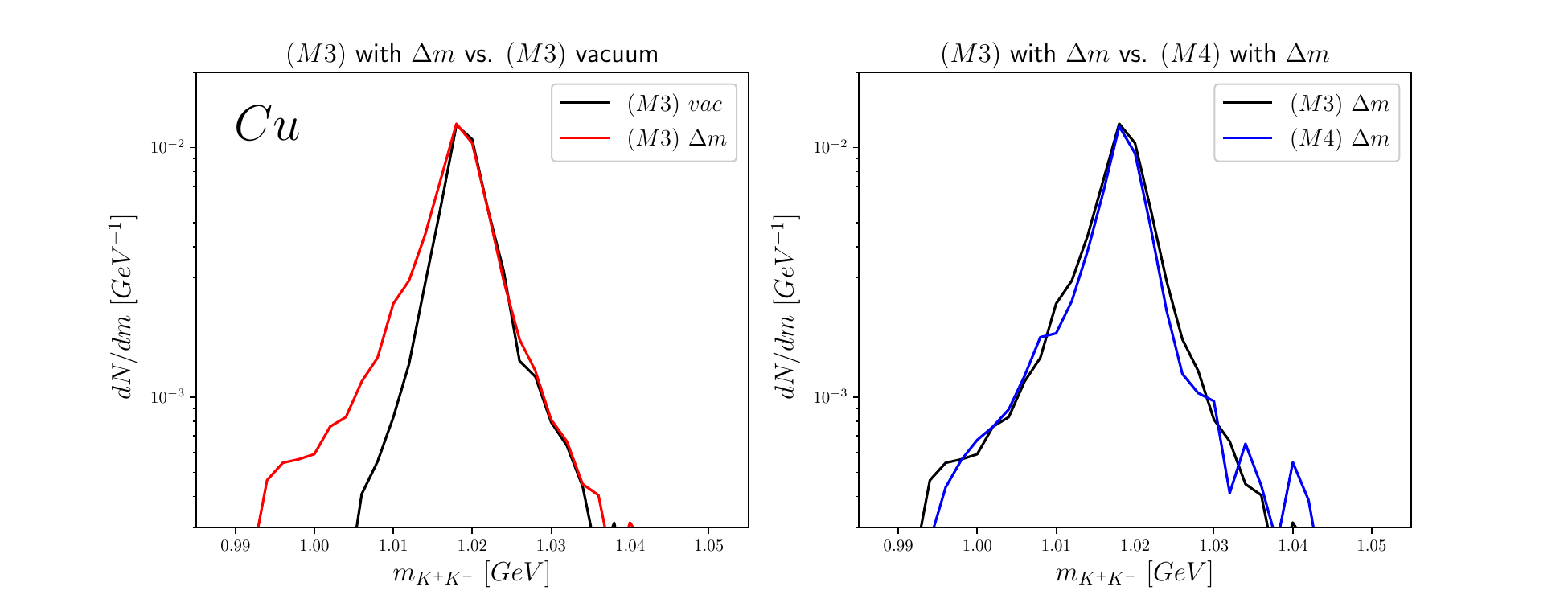}
        \includegraphics[width=4.5in]{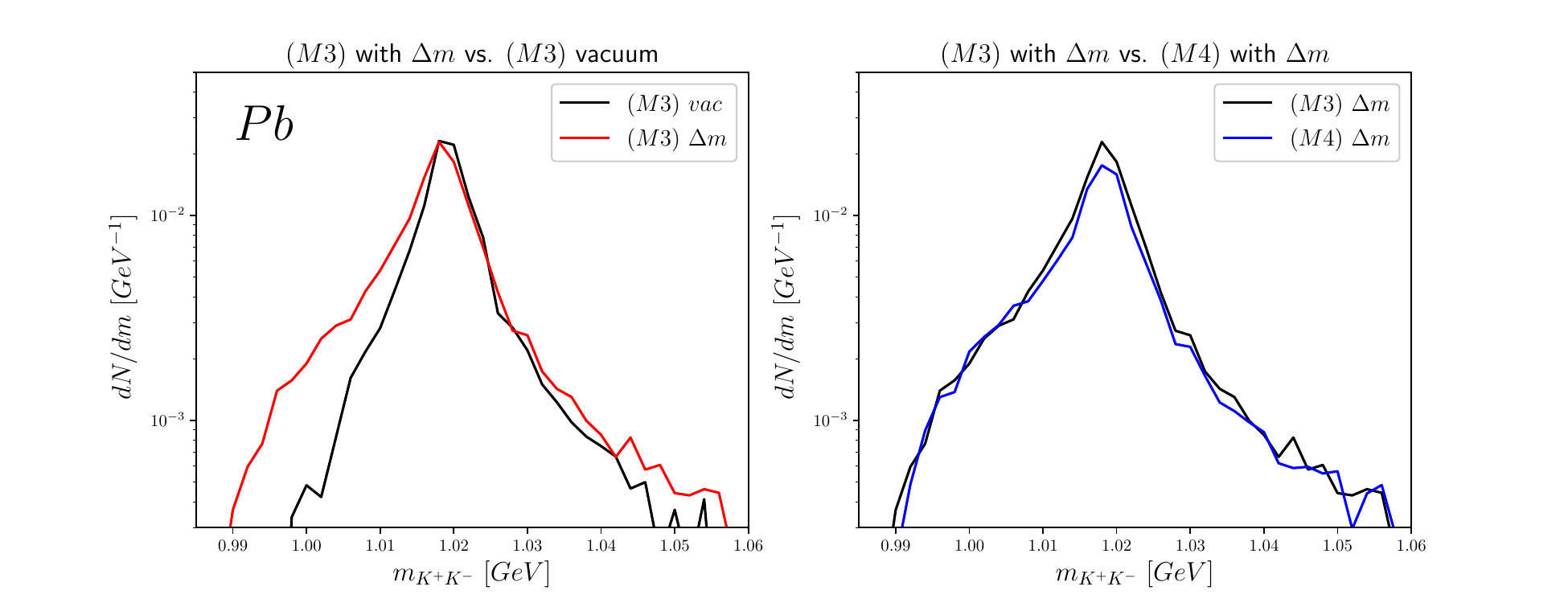}
    \caption{Invariant mass spectra of the K meson pairs when $\Delta m(\rho_0)=-34$ MeV mass shift is applied to the $\phi$ mesons using the (M3) and (M4) mean fields. On the left side, the vacuum (without mean fields, or final state interactions for reference) and the mass shift cases are compared using only the (M3) mean field, while on the right side, the (M3) and (M4) mean field cases are compared using $\Delta m(\rho_0)=-34$ MeV.}
    \label{fig:33}
\end{figure}

After having carefully examined the possible distorting effect due to the mean fields, elastic collisions, and absorptions, we can next turn to identifying the signal of the mass shifts. In the left figures of  Fig.~\ref{fig:33}, a comparison of the applied $\Delta m_0=-34$ MeV mass shift is shown for C, Cu, and Pb targets in 30 GeV proton-induced reactions, where the black curves on the left panels represent the $\Delta m_0=0$ vacuum case, while the red curves show the spectrum generated by including the negative mass shift. All calculations in the left figures have been made with the (M3) potential, including all the relevant final state interactions, without any cuts for the $\phi$ meson momenta. 
The effects of the momentum cuts will be described later at the end of this section.  
As can be seen in the right figures of Fig.~\ref{fig:33}, using the (M4) potential leads to similar results as those obtained with the (M3) potential. 

The shown curves suggest that the mass shift generates an enhancement on the low-mass-side of the peak, which 
could be observed in the region between $1$ and $1.02$ GeV. However, the mass shift effect does not fundamentally alter the shape of the spectrum, making it difficult to identify a clear experimental signal. Therefore, a detailed
analysis with a careful comparison of spectral lineshapes with experimental data will be needed. 

\begin{figure}[h]
    \centering
        \includegraphics[width=4in]{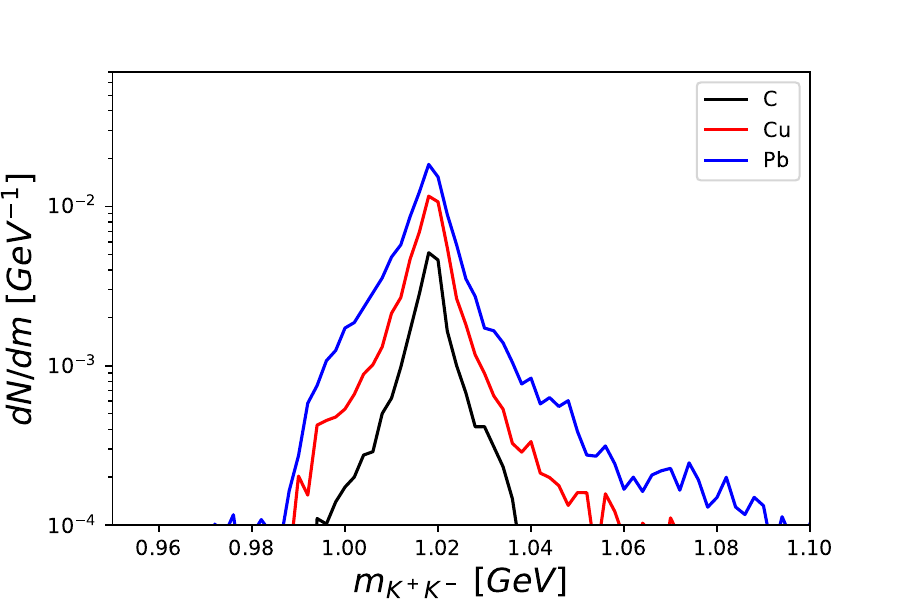}
        \includegraphics[width=4in]{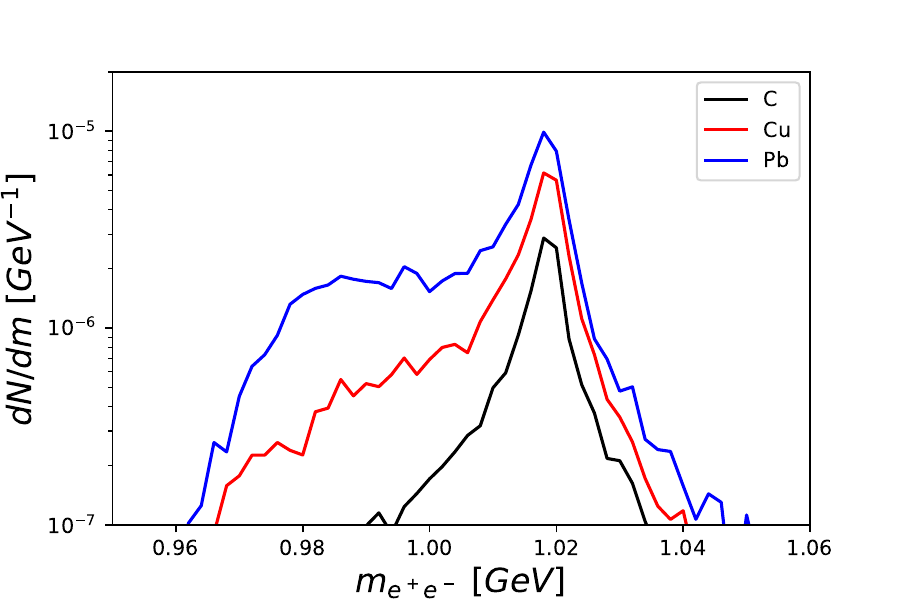}
    \caption{Invariant mass distribution of kaon pairs (upper), and dileptons (lower) for 30 GeV p+C, p+Cu and p+Pb reactions, with a $m(\rho_0)=-34$ MeV mass shift scenario. In these simulations, the (M3) kaon mean fields were applied. }
    \label{fig:34}
\end{figure}

The invariant mass spectrum of kaon pairs is very different from the dilepton case due to the applied mean fields, related collisional and absorption processes and, especially, threshold effects. To see this in Fig~\ref{fig:34}, we compared the invariant mass spectra of the kaon pairs with the dilepton final states, where the difference is apparent, the mass shift having a more significant effect on the enhancement at lower energies in the dilepton case. 
However, it should be noted that its yields are almost 3 orders of magnitude lower than the kaon case, even for Pb targets. Therefore, it will be much more difficult to obtain a statistically significant measurement in the dilepton case. The ideal way to extract the mass shift value would hence be to use both channels and compare them with the results obtained by transport simulations. 

It is interesting to mention in this context that not even the dilepton spectra show a clearly distinguished two-peak structure that was proposed for the $\phi$ in \cite{Ko-Seibert-1994} and was seen in the simulations for the $\Psi(3686)$ charmonium state in central antiproton and proton-induced reactions \cite{Wolf:Charm-2018}. There are two main prerequisites to obtain such a two-peak structure. One is a mass shift much larger than the width of the given particle, the second is a relatively large time spent in the dense region and smaller time in the transition region. In the case of the charmonium studied in \cite{Wolf:Charm-2018} the mass shift was much higher than the width. Furthermore, the time spent in the dense region was much longer than the $\phi$ meson case studied here, because of the same or lower considered bombarding energies, and because the much larger charmonium masses caused the charmonium velocities to be much lower than in the $\phi$ meson. 
One may expect that the two peak structures appear for purely central or near central collisions, which would enable the $\phi$ mesons to decay in a denser region. 

\begin{figure}[!h]
\centering\includegraphics[width=4.5in]{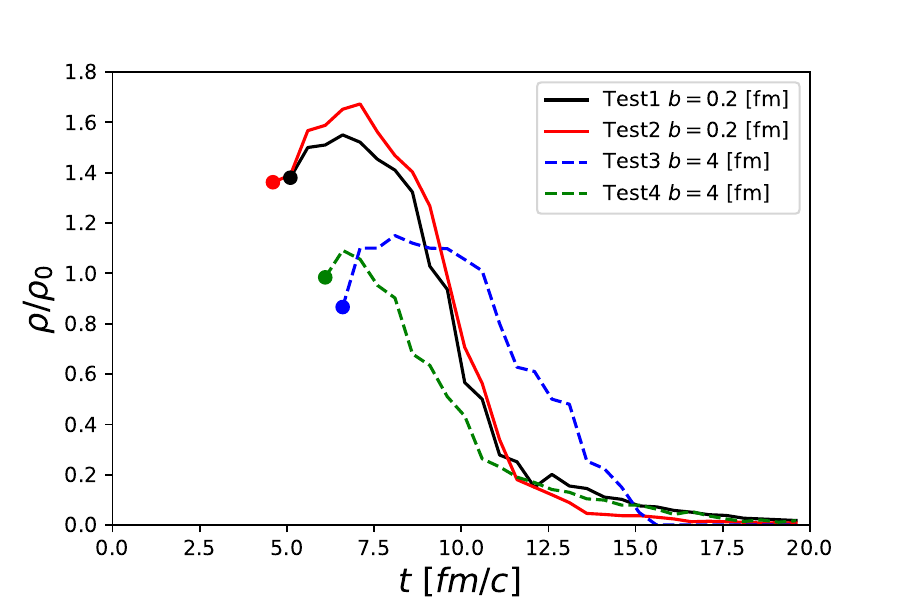}
\caption{Density profile for 2+2 test particles with impact parameters $b=0.2$ fm, and $b=4$ fm. The full circle represents the time, when the specific particle is created, and the full lines correspond to the almost central $b=0.2$ fm case, while the dashed lines correspond to $b=4$ fm collisions.}
\label{fig:35}
\end{figure}

To give more insight into this point, in Fig.~\ref{fig:35}, we compared the density profile that a test $\phi$ particle feels after it is created during its propagation in dense matter at impact parameters $b=0.2$ fm and $b=4$ fm in p+Cu collisions at 30 GeV bombarding energies. In these simulations we have switched off the $\phi$ meson decay, so that we can observe the full density profile the particle feels during the full evolution of the nonequilibrium system. 
There are several points to note from the shown density profiles. Firstly, it is evident that the densities are lower for peripheral collisions, 
which can naturally be expected from the density distribution of the target. 
Another difference to note is the time of particle creation, which tends to be a bit larger for peripheral collisions, which can be easily understood from the geometry of the collision. 
The most significant feature of the different impact parameters, however, is the nature of the transition from the dense region to the vacuum, 
which is rather different in central and peripheral collisions. In the central case, the particles tend to stay in the dense region a longer time ($t \approx 3-9$ fm/c), after which they transition to leave for the vacuum rather quickly ($t \approx 9-12$ fm/c). This allows the $\phi$ mesons to decay in a very dense and approximately constant density region for a longer time than in the transition region, which means that it corresponds to a well-localized, stable, larger contribution in the final invariant mass spectra. 
In contrast, for a larger impact parameter, the maximum density is smaller, and the transition period is longer than in the central case. Due to these effects, the invariant mass spectra will be more smeared out, and hence a distinct second peak will less likely be generated, 
which is exactly the behavior that we have observed in the simulations.
Experimentally, charged particle multiplicity can be used as an approximate measure of the impact parameter, 
which might be useful in the analysis of future experimental data of the J-PARC E16 and E88 experiments. 

\begin{figure}[!h]
\centering\includegraphics[width=4.5in]{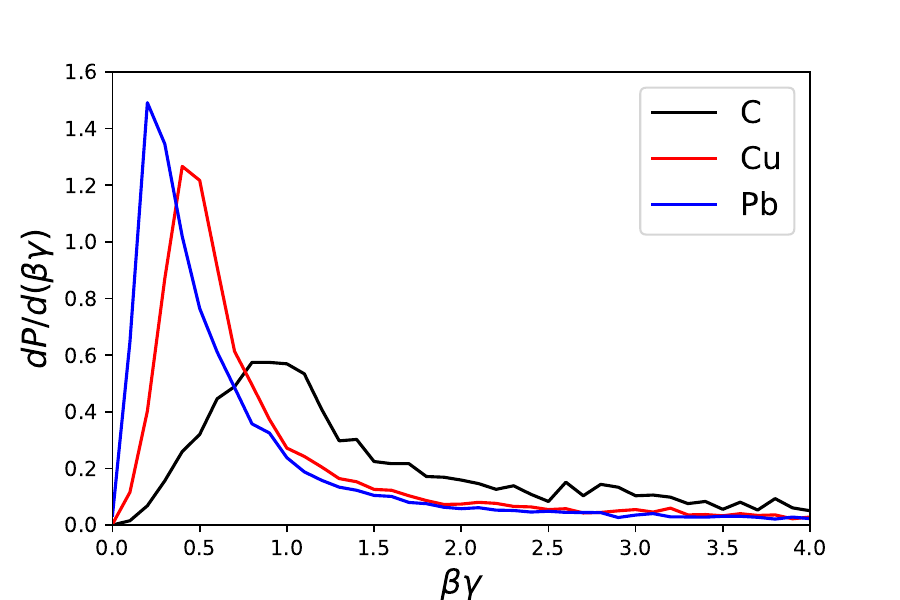}
\caption{The $\beta \gamma$ distributions of $\phi$ mesons in 30 GeV proton+C/Cu/Pb reactions at the time of their decay into $K^+$/$K^-$ pairs.}
\label{fig:36}
\end{figure}

In an experimental setup, cuts that come from, e.g., the geometric acceptance, detector sensitivity, etc., are often directly affecting the measurable quantities. We will here not take such details (for example, of the J-PARC E88 experiment) into account, but will rather compare the results with different $\beta \gamma$ cuts and check whether these alter the possible observability of the mass shifts. 
 In Fig.~\ref{fig:36}, the distribution of the $\beta \gamma$ values of the decaying $\phi$ mesons is shown for $30$ GeV proton-induced reactions using, as before, C, Cu, and Pb targets.

Note that these distributions are taken at the time of the decay of the $\phi$ mesons and not at their time of creation, when their momentum could be much larger. The elastic collisions of $\phi$ mesons with the surrounding nucleons and other hadrons cause their momenta  to 'thermalize' to smaller values, which can be seen in Fig.~\ref{fig:36} by comparing the distributions of the different targets. In the case of a larger target, e.g., Pb, there are more elastic collisions before the $\phi$ decays into kaons. We can therefore expect their momenta to be shifted to lower values by a greater amount than in the case of smaller targets, with fewer elastic collisions. 
From the $\beta \gamma$ distributions shown in Fig.~\ref{fig:36}, it can be seen that a severe cut e.g., $\beta \gamma < 0.5$, could worsen the statistics by a factor of $\int_{0}^{0.5} \frac{dP}{d(\beta \gamma)}  d(\beta \gamma)/\int_{0}^{\infty} \frac{dP}{d(\beta \gamma)}  d(\beta \gamma) \approx 0.05$ in the case of the C target. However, the situation is not that bad for the case of Cu, and Pb, where the corresponding factors for $\beta \gamma<0.5$ are $0.27$, and $0.46$. 
Further detailed studies are needed to clarify the influence of these cuts to the measured invariant mass spectra. A first attempt in this direction is 
discussed in the following paragraph.

\begin{figure}[!h]
\centering\includegraphics[width=4.5in]{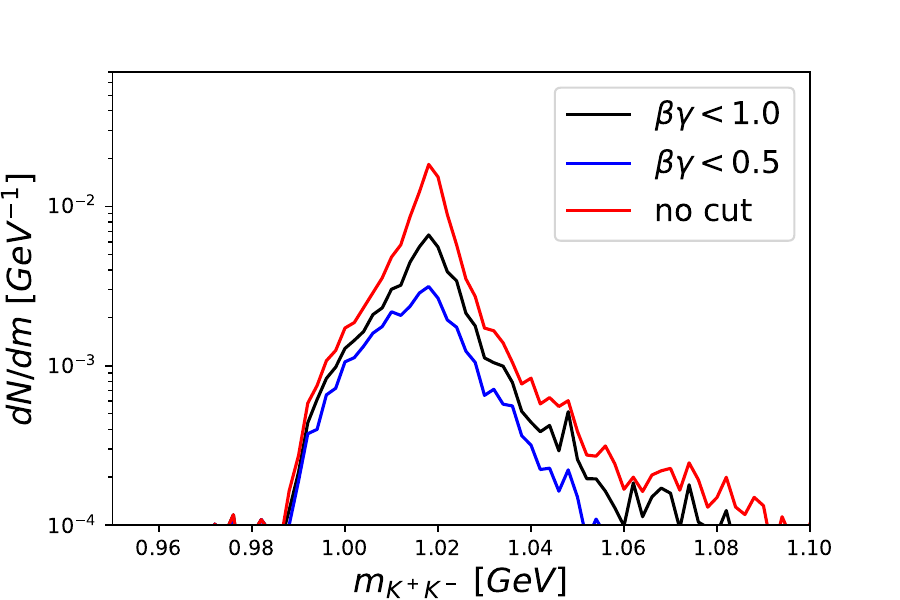}
\caption{The effects of different cuts on the $\phi$ velocity in p+Pb collisions at $30$ GeV bombarding energies using the (M3) mean fields. The red curves shows the case, when no cuts were applied, for reference.}
\label{fig:37}
\end{figure}

The upcoming E16 experiment aims to handle a severe cut of $\beta \gamma < 0.5$. We have hence carried out simulations for the case of two different cuts, to study possible deviations from the results shown in Fig.~\ref{fig:34}. The results for $30$ GeV p+Pb collisions  are shown in Fig.~\ref{fig:37}, where we applied the cuts $\beta \gamma < 0.5$ and $\beta \gamma < 1.0$, and compared the respective spectra to the case with no cuts applied.

From the plot, it can be seen that the main contribution to the peak is significantly reduced by the cuts. While the larger mass regions have the same overall reduction as the vacuum peak, the broadened shoulder at lower invariant masses has a smaller reduction. 
This can be understood from the tendency that particles with smaller $\beta \gamma$ values remain inside of the target for a longer time, thus 
making them more sensitive to, e.g., mass shift effects.
This hence could make it possible to use the cut dependence of the spectra to further constrain the mass shifts at finite density. We will address these issues in our forthcoming studies, in which we will furthermore attempt to provide a prescription to find observables that are most sensitive to the mass shift in an experimental scenario, including a proper uncertainty analysis.

\section{Conclusions}
\label{sec:Conclusions}
In this paper, we have examined the possibility of observing the mass shifts of phi mesons at finite nuclear densities in proton-induced reactions at 30 GeV bombarding energies using C, Cu, and Pb targets. Due to the large branching fractions of the $\phi \rightarrow K^+K^-$ decays, the naive expectation is that, 
as a result of high statistics data, a clear broadened shoulder could be observed in the invariant mass spectra of the final kaon pairs. However, according to our transport simulations, the signal depends heavily on the applied kaonic mean fields and final state interactions between the created kaons and nucleons. We have examined four different scenarios with different mean fields, in which two of them were physically reasonable parametrizations of the repulsive $K^+$ and attractive $K^-$ potentials.  
Our results suggest that, although the invariant mass spectra are strongly influenced by the asymmetry of the $K^+$/$K^-$ mean fields, the final-state interactions between kaons and nucleons tend to smear out this asymmetry in the measured spectra. 
We also find that the non-negligible effects of the mean fields and the final state interactions, make it difficult to observe a clean signal of the mass shift from the invariant mass spectra alone. In contrast to the dilepton spectra, where a cleaner signal is expected (see Fig.\,\ref{fig:34}), especially when using large nuclei (e.g., Pb), the shoulder on the left side of the peak caused by the mass shift is less visible. 
Nevertheless, we observe that the mass shift causes an enhancement that goes beyond the differences due to the different mean fields (see Fig.\,\ref{fig:33}), which can help to constrain the $\phi$ meson mass shift in nuclear matter if precise enough data become available. Furthermore, while the
momentum-integrated spectra are not as clean as in the dilepton case, applying specific cuts to the $\phi$ meson momenta 
can be helpful to obtain data that are more sensitive to
the value of the mass shift near normal nuclear density. A careful analysis of both the final state kaon and the dilepton invariant mass spectra, will 
then hopefully make it possible to specify the nature and degree of spectral modification of the $\phi$ meson in nuclear matter.

On other note, for the kaon decays, depending on the kaon decay angle with respect to the motion of the $\phi$ meson, one can directly isolate the longitudinal and transverse modes of the $\phi$ meson, from which one is able to isolate the momentum dependence that might also smear out the pure mass shift\cite{Park:2022ayr}. 
Such measurement will be possible at the J-PARC E88, which will offer nearly uniform acceptance over kaon emission angles.

Lowering the bombarding energy may also strengthen the effect of a possible mass shift. In this study we investigated the effect of the mass shift at the bombarding energy of 30 GeV, since this is the bombarding energy available at J-PARC. 
However, at lower bombarding energies (5–20 GeV), the produced $\phi$ mesons move significantly slower and therefore remain in the medium for a longer time.  Consequently, the relative contribution of in-medium decays compared to vacuum decays increases for both the dilepton and kaon pair spectra. 
Such energy dependencies can be further studied at the SIS100/FAIR accelerator, where the available bombarding energies are more flexible, and the CBM detector is capable of simultaneously measuring kaons and dileptons. We will investigate the dependence on bombarding energy in a forthcoming paper.
\section*{Acknowledgment}
This work was supported by the Korea National Research Foundation under Grant No.2023R1A2C300302311 and 2023K2A9A1A0609492411, and the Hungarian OTKA fund K138277, and
MEXT/JSPS KAKENHI Grant Numbers JP25H00400 and JP24H00236. 

We thank L. Tolos for providing the data for the M4 potential.

\bibliographystyle{ptephy}
\bibliography{MAIN}
%

\vspace{0.2cm}
\noindent


\let\doi\relax


\end{document}